\documentstyle[12pt]{article}
\begin{document}

\centerline {\Large HIGH FREQUENCY STOCHASTIC RESONANCE}
\centerline {\Large IN PERIODICALLY DRIVEN SYSTEMS}

\vskip 0.5truecm
\centerline {M I Dykman}
\centerline {Department of Physics, Stanford University,}
\centerline {Stanford, CA 94305, USA.}

\vskip 0.5truecm
\centerline {D G Luchinsky\footnote{Permanent address: VNIIMS,
Andreevskaya nab. 2, Moscow 117965, Russia.}}
\centerline {School of Physics and Materials, Lancaster University,}
\centerline {Lancaster, LA1 4YB, United Kingdom.}

\vskip 0.5truecm
\centerline {R Mannella}
\centerline {Dipartimento di Fisica, Universita di Pisa,}
\centerline {Piazza Torricelli 2, 56100 Pisa, Italy.}

\vskip 0.5truecm
\centerline {P V E McClintock, N D Stein and
N G Stocks\footnote{Now at: Department of Engineering, University of Warwick,
Coventry, CV4 7AL, UK.}}
\centerline {School of Physics and Materials, Lancaster University,}
\centerline {Lancaster, LA1 4YB, United Kingdom.}

\vskip 0.75truecm
\noindent
\underbar {Abstract}

High frequency stochastic resonance (SR) phenomena, associated with
fluctuational transitions between coexisting periodic attractors, have been
investigated experimentally in an electronic model of a single-well Duffing
oscillator bistable in a nearly resonant field of frequency $\omega_F$.
It is shown that, with increasing noise intensity, the signal/noise
ratio (SNR) for a signal due to a weak trial force of frequency $\Omega
\sim \omega_F$ at first decreases, then {\it increases}, and finally
decreases again at higher noise intensities: behaviour
similar to that observed previously for conventional
(low frequency) SR in systems with static bistable potentials.  The
stochastic enhancement of the SNR of an additional signal at the
mirror-reflected frequency $\vert \Omega - 2 \omega_F \vert$ is also
observed, in accordance with theoretical predictions.  Relationships
with phenomena in nonlinear optics are discussed.


\newpage
The first decade of research on {\it stochastic resonance} [1] (SR), in which
the signal due to a weak (trial) periodic force in a nonlinear system
can be optimally amplified by the introduction of external noise of
appropriate intensity, has concentrated almost exclusively on systems with
coexisting attractors corresponding to the minima of a symmetrical, static,
bistable potential.  In such cases, the noise-induced amplification arises
through the occurrence of fluctuational transitions between the attractors.
For suitably chosen noise intensity, these become nearly periodic at the
frequency of the trial force, with an average amplitude that can approach
the half-separation of the attractors.  This is the mechanism responsible
for SR phenomena studied in connection with ice-ages, ring-lasers,
electronic circuits, passive optical systems, electron spin
resonance, sensory neurons, a magnetoelastic ribbon, and a laser with
saturable absorber; we shall refer to it as
conventional SR.

More recently, efforts have been initiated to seek SR in other classes of
systems having quite different kinds of attractors: it has now been
identified in underdamped nonlinear oscillators that have single static
attractors [2], in a bistable, chaotic, electrical circuit with two
coexisting attractors of which one is a limit cycle and the other is
chaotic [3], and in a system with two coexisting stable limit cycles that
have the {\it same period} and correspond to periodically forced vibrations
of a damped nonlinear oscillator [4]; the onset of a resonant absorption
that increased extremely sharply with noise intensity - a phenomenon that
has much in common with SR - had earlier been demonstrated theoretically [5]
for the latter system.  In this Letter we shall show that SR
of this latter kind, while similar to conventional SR in some respects,
also has a number of interesting features that distinguish it from
earlier manifestations of the phenomenon and which have implications for
four-wave mixing in nonlinear optics.  We expect the ideas to be applicable
to a large class of passive optically bistable systems and, in particular,
to optically bistable microcavities [6].

The system that we consider is the nearly-resonantly-driven,
underdamped, single-well Duffing oscillator with additive noise,

\begin{equation}
\ddot q + 2 \Gamma \dot q + \omega^2_0 q + \gamma q^3 = F \cos (\omega_F t)
+ f (t)
\end{equation}

$$\Gamma, \vert \delta \omega \vert \ll \omega_F, \qquad \gamma \delta
\omega > 0,
\qquad \delta \omega = \omega_F - \omega_0$$

$$\langle f (t) \rangle = 0, \qquad \langle f(t) f (t^{\prime})\rangle =
4\Gamma T \delta (t - t^{\prime})$$

\noindent
This system is of topical interest on account of its importance in
nonlinear optics [7] and its relevance to experiments on a confined
relativistic electron excited by cyclotron resonant radiation [8].  Provided
that $F^2 \ll \omega_0^4 (\delta \omega^2 + \Gamma^2)/\vert \gamma \vert$,
and that the noise is weak, the resultant comparatively small amplitude
$[\ll (\omega_0^2/\vert \gamma \vert)^{\frac {1}{2}}]$ oscillations of $q (t)$
can conveniently be discussed in terms of the dimensionless parameters [5]

\begin{equation}
\eta = \Gamma /\vert \delta \omega \vert, \qquad \beta = \frac {3 \vert
\gamma \vert F^2}{32 \omega^3_F (\vert \delta \omega \vert)^3}, \quad \alpha
= 3 \vert \gamma \vert T/8\omega^3_F \Gamma
\end{equation}

\noindent
which characterise, respectively, the frequency detuning, the strength
of the main periodic field, and the noise intensity.  The bistability
[9] in which we are interested arises for a restricted range of $\eta$ and
$\beta$: within the triangular region bounded by the full lines of Fig 1. Thus,
as the amplitude of the periodic force is gradually increased from a small
value at fixed frequency (see vertical line a - a$^{\prime}$),
the system moves from monostability (one small limit cycle), to bistability
(two possible limit cycles of different radii), and then back again to
monostability (one large limit cycle).  The effect [5] of weak noise $f (t)$
is to cause small vibrations about the attractors, and to induce occasional
transitions (cf [10]) between them when the system is within the bistable
regime.  We shall see that SR phenomena occur in the close vicinity
of the kinetic phase transition (KPT) line [11], shown dashed in Fig 1, where
the populations of the two attractors are equal.

Our principal aim is to consider the response of the system (1) to an
additional weak trial force $A \cos (\Omega t + \phi)$. The combined effects
of dissipation and noise result in a steady statistical distribution, and
the response of the system can therefore be described, in terms of linear
response theory, by a susceptibility.  The trial force beats with the main
periodic force and thus gives rise to vibrations of the system,
not only at $\Omega$, but also at the combination frequencies $\vert \Omega
\pm 2n \omega_F \vert$ (and also at $\vert \Omega \pm (2n+1) \omega_F \vert$ in
the case of nonlinearity of a general type).  We are interested in
the case where the strong and trial forces are
both nearly resonant: that is, $\omega_F$ and $\Omega$ both lie close to
$\omega_0$.  This is the case for which the response to the trial force is
strongest.  It is at its most pronounced at frequency $\Omega$ and at the
nearest resonant combination, which for (1) is $\vert \Omega - 2\omega_F
\vert$.  The
amplitudes of vibrations at these frequencies can be described respectively by
susceptibilities $\chi (\Omega )$, $X (\Omega )$, so that the
trial-force-induced modification of the coordinate {\it q}, averaged over
noise, can be sought in the form

\begin{equation}
\delta \langle q (t) \rangle = A {\rm Re} \{ {\cal \chi} (\Omega ) \exp [-i
\Omega t - i \phi] + X (\Omega ) \exp [i (2 \omega_F - \Omega ) t] - i\phi]\}
\end{equation}

\noindent
Within the KPT range, for
$\Omega$ close to $\omega_F$, $\vert {\rm Im} \chi (\Omega) \vert$ becomes
large and strongly noise-dependent [5].  The rapid rise in susceptibility with
noise intensity corresponds precisely to SR since, according to (3), the
squared amplitudes of the signals at frequencies $\Omega$ and $\vert \Omega
- 2 \omega_F \vert$ (and the integrated powers of the corresponding peaks
in the power spectrum) are

\begin{equation}
S (\Omega ) = \frac {1}{4} A^2 \vert \chi (\Omega ) \vert^2, \quad S
(\vert \Omega - 2 \omega_F \vert ) = \frac {1}{4} A^2 \vert X (\Omega ) \vert^2
\end{equation}

\noindent
An intuitive understanding of the mechanism of stochastic amplification
can be gained by noting that the trial force modulates the driving force
(and the coordinate $q(t)$) at frequency $\vert \Omega - \omega_F \vert$
and that, when $\vert \Omega - \omega_F \vert$ is
small, the system responds almost adiabatically.  In
terms of the phase diagram Fig 1, the beat envelope
then results in a slow vertical oscillation of the operating point.
If the operating point is taken to be p, and its range of modulation
${\rm p}^{\prime}-{\rm p}^{\prime \prime}$ is set to straddle the KPT line
as shown, and the noise intensity is optimally chosen, then the
system will have a tendency to make inter-attractor transitions
{\it coherently}, once per half-cycle of the beat frequency.  The net effect
of the noise is therefore
to increase the modulation depth of the beat envelope of the response,
thereby increasing the components of the signal at frequencies $\Omega$
and $\vert \Omega - 2\omega_F \vert$.

We are investigating the response of the system (1), and the variation of
the signal/noise ratio with $\alpha$, through analogue experiments on
an electronic model of (1), of which the relevant technical details were
given in a recent conference report [4].
In terms of scaled units the circuit parameters were set, typically, to:
2$\Gamma$ = 0.0397; $\omega_0$ = 1.00; $\gamma$ = 0.1; $\omega_F$ = 1.07200;
$\Omega$ = 1.07097; and, to seek SR near the KPT, {\it F} = 0.068 and the
amplitude of the trial force {\it A} = 0.006.    A
spectral density of fluctuations of the coordinate $q(t)$ (about
$\langle q(t) \rangle$ for $A$=0)recorded with the above parameter values
for $\alpha$ = 0.061 and 16384 samples in each
realization, is shown in Fig 2.  The smooth background is the supernarrow
peak of [11], here broadened by noise (although its width remains very much
smaller than $2 \Gamma$); delta function spikes, indicated by
raised points [12] of the discrete spectrum, are clearly visible, not only at
the trial force frequency ($\Omega$), but also as predicted at the
mirror-reflected frequency $(2 \omega_F - \Omega)$.

The signal/noise ratios $R$, determined in the usual way [1] from measurements
of the delta spikes and the smooth background, are plotted (data points) as
functions of noise intensity $\alpha$ in Fig 3 for $\beta$ =
0.814, $\eta$ = 0.236.  It is immediately evident that there is a
range of $\alpha$ within which $R$ {\it increases} with $\alpha$.
It is also apparent that, for both the main and
the mirror-reflected signals, the form of $R (\alpha)$ in Fig 3 is
remarkably similar to that observed earlier [1] in the case of
conventional SR.  That is, $R$ initially decreases with $\alpha$,
on account of the increase in its denominator.
With further increase of $\alpha$, the inter-attractor transitions come
into play and become phase-coherent
with the trial force to a high probability, giving rise to an increase in
$R$ through the stochastic amplification mechanism discussed above.
Finally, for still larger $\alpha$, $R$ decreases again partly owing to
the continuing rise in its denominator and partly because transitions
are then occurring very frequently, within individual periods of the trial
force, with a corresponding reduction in the proportion of the phase-coherent
jumps that are responsible for the amplification.

A quantitative theory of the phenomenon is readily constructed through
an extension [13] of [5].
It leads to contributions to the susceptibilities from
inter-attractor transitions of the form

\begin{equation}
\chi_{tr} (\Omega ) = \frac {w_1 w_2}{2\omega_F (\omega_F - \omega_0)}
(u_1^* - u_2^*) \frac {\mu_1 - \mu_2}{\alpha} \left[ 1 -
\frac {i (\Omega - \omega_F)}{W_{12} + W_{21}}\right]^{-1}
\end{equation}

$$X_{tr} (\Omega ) = \frac {u_1 - u_2}{u^*_1 - u_2^*} \chi_{tr} (\Omega), \quad
\mu_j = \sqrt {\beta} (\frac {\partial R_j}{\partial \beta})$$

\noindent
where $w_1, w_2$ are the populations of the attractors 1, 2 and
$W_{12}, W_{21}$ are the probabilities of transitions between them, which
are of the activation type $W_{ij} \propto \exp (-R_i/\alpha)$.
(The $u_i, u^*_i$, which define the positions of the attractors in the
rotating coordinate frame, can be regarded as constants for given
$\eta, \beta$).  It is evident that the contributions (5)
come into play if, and only if, the system is within the KPT range
where the populations of the unperturbed attractors are comparable:
otherwise, the factor $w_1 w_2 \propto \exp (-\vert R_1 - R_2 \vert/\alpha)$
will be exponentially small.  Within the KPT range, however, the
susceptibilities will be large because they are proportional to the large
factor $\vert \mu_1 - \mu_2 \vert/\alpha$; the rapid increase of $W_{ij}$ with
noise intensity then ensures that there will be a range of $\alpha$ in which
both susceptibilities increase very rapidly with $\alpha$, consistent with the
experiments.  The full theory [13], including the effect of intra-attractor
vibrations, leads to the curves of Fig 3.  Given the large
systematic errors inherent in the measurements - arising e.g. from
$\delta \omega$ (1), a small difference between large quantities which, in
$\beta$ (2), is then raised to its third power - the agreement between theory
and experiment can be considered excellent.

In conclusion, we would emphasize, first, that, in contrast to earlier
forms of bistable SR [1, 3], stochastic amplification occurs here
{\it not} at the relatively low frequency of the quasi-periodic
inter-attractor hopping but, rather, at $\Omega$ close to the much higher
(tunable) frequency $\omega_F$ of the main periodic driving force.
To emphasize the distinction,
it seems appropriate to refer to the new form of SR as {\it high frequency
stochastic resonance} (HFSR).

Secondly, we draw attention to the relationship of HFSR to four-wave mixing in
nonlinear optics [14]. In effect, our results correspond to noise-enhanced
amplification of the signal wave, and noise-enhanced generation of the idler
wave.  The mechanisms are  {\it resonant} and, although they have the
appearance of four-wave mixing, they actually correspond to multiple-wave
processes: in terms of quantum optics, the oscillator absorbs and re-emits
many quanta of the strong field.  The very high
amplification/generation coefficients arise partly from their resonant
character and partly from the fact that the signal sizes correspond, not to
the amplitudes of vibrations about the attractors but, as usual in bistable
SR, to the \lq \lq distance" between the attractors (to their coordinate
separation
for conventional SR, and approximately to their difference in amplitude
in the present case).

Finally, our prediction and demonstration of HFSR for periodic attractors,
and its similarity (Fig 3) to conventional SR, leads to a broader
and more general perception of the physical nature of bistable SR.
Like the onset of supernarrow peaks in the power spectra, conventional SR
[1] and HFSR are both critical phenomena that arise in the range of the KPT.
While HFSR is to be anticipated for coexisting stable limit cycles with
equal periods, low-frequency SR is a more robust effect.  It arises, not only
for systems fluctuating in simple double-well potentials but also for systems
where one (or both) of the attractors is chaotic [3]; we may infer from [5]
that, in general, low frequency SR is also to be anticipated for periodic
attractors [although not for (1), where the centres of the forced
vibrations are independent of amplitude].  Since noise gives rise to
fluctuational hopping between any type(s) of attractors, it seems reasonable
to conclude that SR is actually a quite general phenomenon characteristic of
{\it all} systems with coexisting attractors, regardless of the nature of
those attractors, provided only that the system lies within its KPT range.

This research has been supported by the Science and Engineering Research
Council (United Kingdom), by the Royal Society of London and by the
European Community.

\newpage
\centerline {References}

\begin{itemize}
\item[1.] For a recent review, see {\em J. Stat. Phys.} {\bf 70}, nos 1/2
 (1993), special issue on stochastic resonance.

\item[2.] N G Stocks, N D Stein, S M Soskin and P V E McClintock,
{\it J Phys A} {\bf 25} L1119 (1992); N G Stocks, N D Stein and
P V E McClintock, {\it J Phys A} {\bf 26} L385 (1993).

\item[3.] V S Anishchenko, M A Safonova and L O Chua,
{\it Int. J. of Bifurcation and Chaos} {\bf 2}, 397 (1992).

\item[4.] M I Dykman, D G Luchinsky, R Mannella, P V E McClintock,
N D Stein and N G Stocks, {\it J. Stat. Phys}. {\bf 70}, 479 (1993).

\item[5.] M I Dykman and M A Krivoglaz, {\it Sov. Phys. J.E.T.P.} {\bf 50},
30 (1979).

\item[6.] H J Carmichael in {\em Optical Instabilities}, ed. R W Boyd,
 M G Raymer and L M Narducci, Cambridge University Press, 1986, p111.

\item[7.] H M Gibbs, {\it Optical Bistability: Controlling Light with Light},
Academic Press, New York, 1985; P D Drummond and D F Walls, {\it J. Phys. A}
{\bf 13} 725 (1980); Chr. Flytzanis and C L Tang, {\it Phys. Rev. Lett.}
{\bf 45}, 441 (1980); J A Goldstone and E Garmire, {\it Phys. Rev. Lett.}
{\bf 53}, 910 (1984).

\item[8.] G Gabrielse, H Dehmelt and W Kells, {\it Phys. Rev. Lett.}
{\bf 54}, 537 (1985).

\item[9.] L D Landau and E M Lifshitz, {\it Mechanics} (Pergamon, London,
1976).

\item[10.] M R Beasley, D D'Humieres and B A Huberman, {\it Phys. Rev. Lett.}
{\bf 50}, 1328 (1983).

\item[11.] M I Dykman, R Mannella, P V E McClintock and N G Stocks,
{\it Phys. Rev. Lett.} {\bf 65}, 48 (1990).

\item[12.] The value of $\Omega$, the number of points in each $q (t)$
realization, and the sample interval, were chosen to be such that the signals
at $\Omega$ and $(2 \omega_F - \Omega)$ each fell within individual bins of
the discrete SDF, and such that there were several bins in between them.

\item[13.] M I Dykman, D G Luchinsky, R Mannella, P V E McClintock, N D Stein
and N G Stocks, in preparation.

\item[14.] Y R Shen, {\it The Principles of Nonlinear Optics} (Wiley,
New York, 1984).
\end{itemize}

\newpage
\centerline {Figure Captions}

\begin{itemize}
\item[1.] Phase diagram for (1) in terms of reduced parameters (2): the
cuts a-a$^{\prime}$, p$^{\prime}$-p-p$^{\prime \prime}$ are discussed
in the text.

\item[2.] Power spectral spectral density $Q (\omega)$ measured for the
electronic model, with the contents of each FFT memory address shown as a
separate data point on a highly expanded abscissa; a smooth curve, peaking
at $\omega_F$, is drawn through the background spectrum;
vertical lines indicate the delta spikes resulting from the trial force.

\item[3.] The signal/noise ratio $R$ of the response of the system (1) to a
weak
trial force at frequency $\Omega$, as a function of noise intensity $\alpha$,
in experiment and theory: at the trial
frequency $\Omega$ (circle data and associated theoretical curve); and at the
\lq\lq mirror-reflected" frequency ($2 \omega_F - \Omega$) (squares).  For
noise intensities near those of the maxima in $R( \alpha)$, the asymptotic
theory is only qualitative and so the curves are shown dotted.
\end{itemize}
\end{document}